\documentclass[prd,twocolumn]{revtex4}
\pdfoutput=1
\usepackage{amssymb,latexsym}
\usepackage{amsmath,amsbsy,bbm}
\usepackage{ifpdf}
\usepackage{epsfig,bm}
\usepackage{graphicx,comment}
\usepackage{color}
\usepackage{soul}
\usepackage{mathtools}
\usepackage{comment}
\usepackage[normalem]{ulem}
\begin{document}

\title{Detection of Berezinskii--Kosterlitz--Thouless transitions for the two-dimensional
  $q$-state clock models with neural networks}
\author{Yaun-Heng Tseng}
\affiliation{Department of Physics, National Taiwan Normal University,
	88, Sec.4, Ting-Chou Rd., Taipei 116, Taiwan}
\author{
	Fu-Jiun Jiang}
\email[]{fjjiang@ntnu.edu.tw}
\affiliation{Department of Physics, National Taiwan Normal University,
	88, Sec.4, Ting-Chou Rd., Taipei 116, Taiwan}

\begin{abstract}
  Using the technique of supervised neural networks (NN), we study the phase
  transitions of two-dimensional (2D) 6- and 8-state clock models on the square
  lattice. The employed NN has only one input layer, one hidden layer of 2 neurons, and one output layer. In addition, the NN is trained
  without any prior information
  of the considered models. Interestingly, despite its simple architecture, the
  built supervised NN not only detects both the two Berezinskii--Kosterlitz--Thouless (BKT)
  transitions, but also determines the transition
  temperatures with reasonable high accuracy. It is remarkable that a NN, which has extremely simple structure and
  is trained without any input of the studied models, can be employed to study topological
  phase transitions. The outcomes shown here as well as those previously demonstrated
  in the literature suggest the feasibility of constructing a universal NN
  that is applicable to investigate the phase transitions of many systems.

\end{abstract}

\maketitle

\section{Introduction}\vskip-0.3cm

When phase transitions are concerned, apart from the well-known first- and second-order phase transitions
which are related to spontaneous symmetry breaking,
there is a novel type of phase transitions associated with topological defects \cite{Ber71,Ber72,Kos72,Kos73,Kos74}.
Unlike the first- and second
order phase transitions which are characterized by the behavior of the so-called order parameters, this kind of
novel phase transitions, namely the Berezinskii--Kosterlitz--Thouless (BKT) transitions, cannot be understood quantitatively by any order parameter. 
Two-dimensional (2D) $XY$ model on the square lattice is a typical model exhibiting the BKT transition.

The 2D $q$-state clock models are simplified version of the 2D $XY$ model,
Instead of continuous spin values like those of the $XY$ spin, the clock spins are discrete.
These models have very rich phase structures, hence have been studied extensively in the
literature \cite{Jos77,Car80,Roo81,Tob82,Tom01,Tom021,Tom022,Sur05,Lap06,Bae09,Bae10,Ray12,Ort12,Kum13,Cha18,Sur19}.
For $q \le 4$, the $q$-state clock models exhibit one second-order (Ising-type) phase transition from the ferromagnetic phase to the
disordered phase.
For $q \ge 5$, the models have two BKT-type transitions: one from the long-range order phase (LRO) to the pseudo-long-range order phase (PLRO,
and the related transition temperature has a symbol of $T^2_c$)
and the other from PLRO to the paramagnetic phase (The associated critical temperature is denoted as $T^1_c$). The 2D $XY$ model is recovered
from the $q$-state clock model when $q \rightarrow \infty$. It is well-established that the values of $T^1_c \sim 0.892$ do not change
appreciably with $q$.

Recently, techniques of Machine learning (ML) have been applied to many fields of physics \cite{Rup12,Sny12,Baldi:2014pta,Mnih:2015jgp,Oht16,Hoyle:2015yha,Car16,Nie16,Den17,Hu17,Li18,Chn18,Lu18,But18,Sha18,Rod19,Zha19,Tan20.1,Tan20.2,Larkoski:2017jix,Aad:2020cws,Nicoli:2020njz,Miy21,Tan21,Tse22}. In particular, neural networks (NN) are considered
to classify various phases of many-body systems. Both supervised and unsupervised NN have been demonstrated to be able to determine the
critical points of phase transitions accurately for numerous models \cite{Car16,Nie16,Li18,Tan21}.

The conventional NNs known in the literature have very complicated architectures. Typically these NNs have various layers and each layer has many independent nodes (neurons). Moreover,
the associated trainings use real physical quantities, such as the spin configurations or the correlation functions, as the training sets. 
As a result, conducting studies with these conventional NNs demands huge amount of computer memory and is very time consuming. In particular,
the investigations are limited to small to intermediate system sizes. Because of this, the detection of topological phase transitions with the
methods of NN is more challenging when compared with the phase transitions related to spontaneous symmetry breaking. 

Unlike the conventional NNs, extremely simple supervised and unsupervised NNs consisting of one input layer, one hidden layer, and one output layer
are constructed in Refs.~\cite{Tan21,Tse22,Pen22}. In addition, the trainings for these unconventional NNs use no information of the studied systems.
Instead, two artificially made one-dimensional (1D) configurations are employed as the training sets. As a result, the associated
training procedure is easier to implement and is
much more efficient than the conventional approaches. We would like to empahsize the fact that no
inputs of the considered systems, such as the vortex configurations, the histograms of spin orientations, the spin correlation functions, and
the raw spin configurations, are used for training these unconvetional NNs. 
It is demonstrated that the NNs resulting from
these unusual training strategies are very efficient. In other words, it takes much less time to conduct the associated NN calculations.
Particularly, these unconventional NNs can be considered to detect the phase transitions of many three-dimensional
(3D) and two-dimensional (2D) models. Finally, there is also no system size restriction for these unconventional NNs as well and they can be
recycled to study the phase transitions of other systems not considered in Refs.~\cite{Tan21,Tse22,Pen22}.

Between the conventional supervised and unsupervised NNs, unsupervised ones are preferred when the detection of phase transitions are considered.
This is because no prior information of critical point of the studied system is needed when one carries out an unsupervised investigation. In other words, less
efforts in preparation is required when an unsupervised study is performed.

For the mentioned unconventional supervised and unsupervised NNs, the trainings are conducted without any prior information or input from
the considered models. Consequently, supervised one will be a better choice since it takes less time to complete the associated training and
prediction processes. Due to this fact, in this study we directly adopt the supervised NN of Ref.~\cite{Pen22} to study the phase transitions of 2D
6- and 8-state clock models.

Interestingly, the simple supervised NN employed here not only detect both the BKT-type transitions of the 6- and 8-state clock models,
it also estimates the transition temperatures with reasonable good accuracy. It is remarkable that a NN trained without any input of
the considered systems can successfully map out the non-trivial topological phase structures of these studied models. Similar to the
unsupervised NN considered in Ref.~\cite{Tse22}, it is anticipated that the simple supervised NN used here can be directly applied
to study the phase transitions of other models, such as the three-dimensional (3D) $O(3)$ model, the 2D generalized $XY$ model,
the one-dimensional (1D) Bose--Hubbard model, and the 2D $q$-state ferromagnetic Potts model, without carrying out any re-training.

The rest of the paper is organized as follows. After the introduction, the considered models and the built supervised NN are described
in Secs.~II and III, respectively. Then we present the NN outcomes in Sec.~IV. In particular, the critical temperatures of the studied
phase transitions are determined with good precision. Finally, we conclude our investigation in Sec.~V.

\begin{figure*}
	\includegraphics[width=0.8\textwidth]{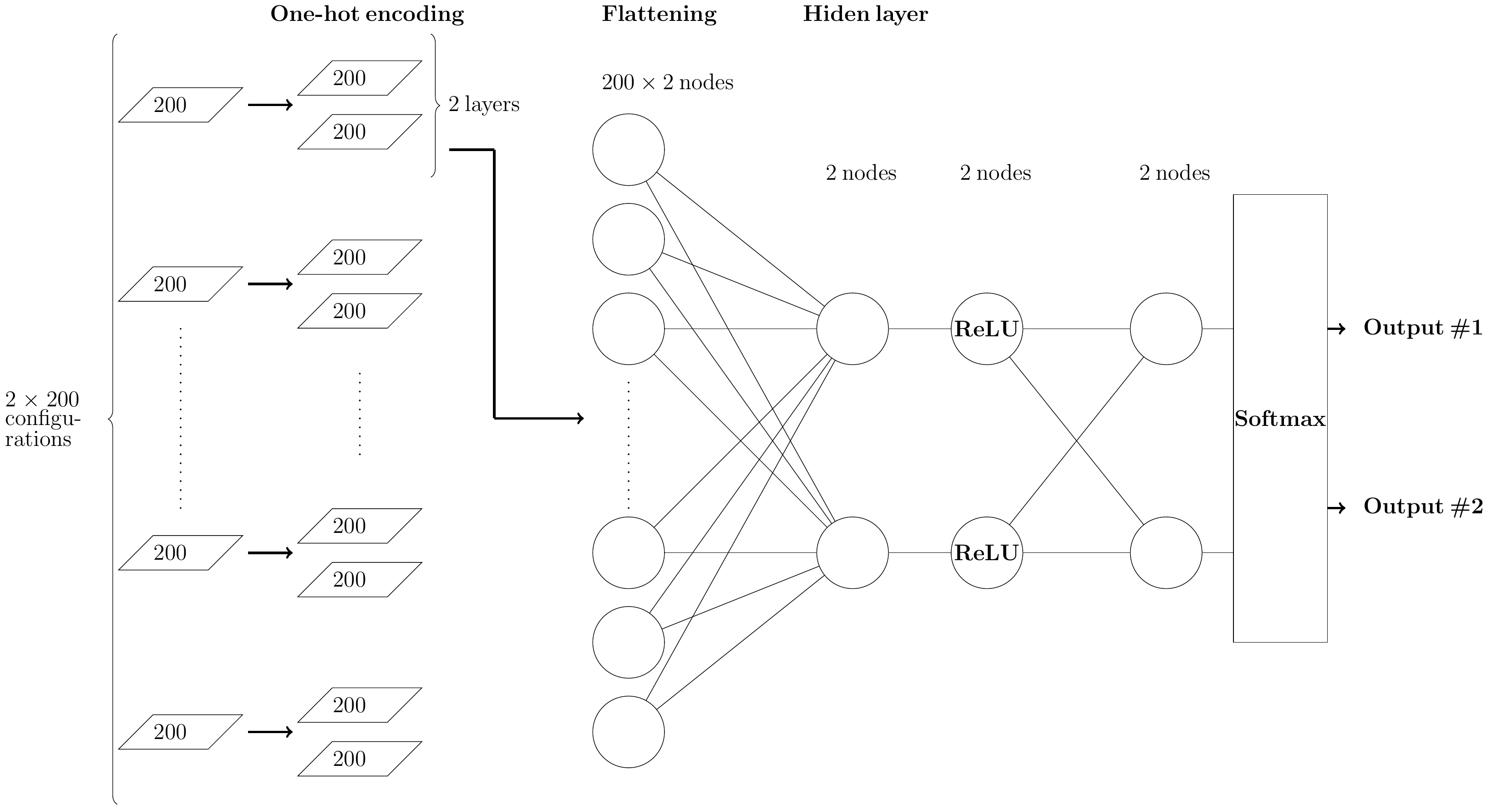}        
	\vskip-0.2cm
	\caption{The MLP employed in this study.}
	\label{figMLP}
\end{figure*}

\section{The considered models}

The Hamiltonian of the 2D $q$-state clock model on the square lattice considered here has the following expression \cite{Cha18}
\begin{equation}
H = -\sum_{\left< ij\right>} \vec{\sigma}_i\cdot\vec{\sigma}_j,
\label{eqn}
\end{equation}
where $\left<ij\right>$ refers to nearest
neighbor sites $i$ and $j$, and $\vec{\sigma}_i$ is a vector at site $i$ with
$\vec{\sigma}_i = \left(\cos\theta_i,\sin\theta_i\right)$.
Here $\theta_i = \frac{2\pi k}{q}$ with $k = 1,2,...,q-2,q-1$.

\section{The constructed NN}

The supervised NN, namely a multilayer perceptron (MLP)
is built using the keras and tensorflow \cite{kera,tens}. In addition, it has extremely simple
architecture. Specifically, the constructed NN has only one input layer, one hidden layer
of two neurons, and one output layer. The considered algorithm and optimizer are minibatch and
adam (learning rate is set to 0.05), respectively.
The activation function ReLU (softmax) is applied in the hidden (output) layer.
The definitions of ReLU and softmax are given by

\begin{eqnarray}
&&\text{ReLU}(x) = \text{max}(0,x),\\
&&\left(\text{softmax}(x)\right)_i = \frac{e^{x_i}}{\sum_{j}e^{x_j}}.   
\end{eqnarray}

One-hot encoding, flattening, and $L_2$ regularization are used as well. Finally, the loss function considered is the
categorical crossentropy $C$ which is defined as

\begin{equation}
C = -\frac{1}{n}\sum_x\sum_j^2 y_j\ln a_j,  
\end{equation}
where $n$ is the number of objects included
in each batch and $a_j$ are the outcomes obtained after applying all
the layers. Moreover, $x$ and $y$ are the training inputs and the corresponding designed labels,
respectively.

Figure \ref{figMLP} is adopted from Ref.~\cite{Pen22} and is the cartoon representation of the employed supervised NN.

The training of the supervised NN is conducted using 200 copies of two one-dimensional artificially made configurations
(Each of which consists of 200 sites) as the training set. These configurations contain no information from the considered models.
Specifically, one (the other) configuration has 1 (0) as the values for all of its elements.
Due to the used training set, the labels employed are two-component vectors (0,1) and (1,0).
On a server with two opteron 6344 and 96G memory, the training takes only 24 seconds.

\section{Numerical Results}

\subsection{Preparation of configurations for the NN prediction}

Using the Wolff algorithm \cite{Wol89}, we have generated several thousand configurations for the 6- and 8-state clock models
with various temperatures $T$ and linear system sizes $L$. For each produced clock configuration, the angles $\theta$ of
2000 sites, which are randomly chosen, are stored. From these stored variables, two hundred are picked randomly and the
resulting $\theta\,\, \text{mod}\,\, \pi$ are used
to build a 1D configuration consisiting of 200 sites which will then be employed for the NN prediction. 

\subsection{The NN results associated with 6-state clock model}

\begin{figure}
       \includegraphics[width=0.4\textwidth]{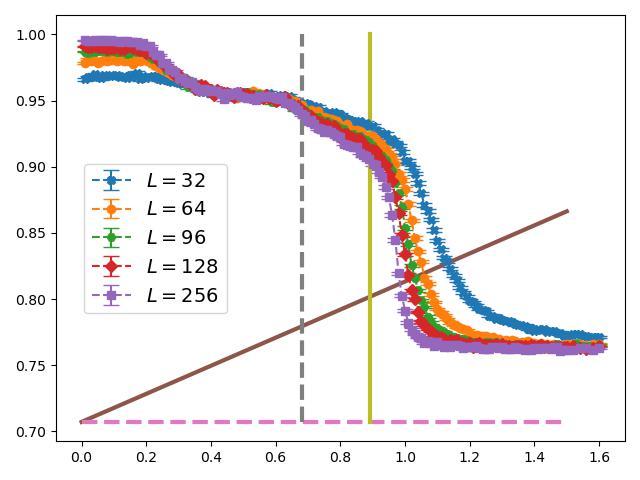}        
        \vskip-0.2cm
        \caption{$R$ as functions of $T$ for various $L$ for the 6-state clock model.
          The dashed horizontal line is 0.70712. The vertical dashed and solid lines
          are the expected transition temperatures $T^{2}_c$ and $T^1_c$. The tilted line is $2 T/(\pi q) + 0.70712$ with $q=6$.}
        \label{fig61}
\end{figure}

\begin{figure}
       \includegraphics[width=0.4\textwidth]{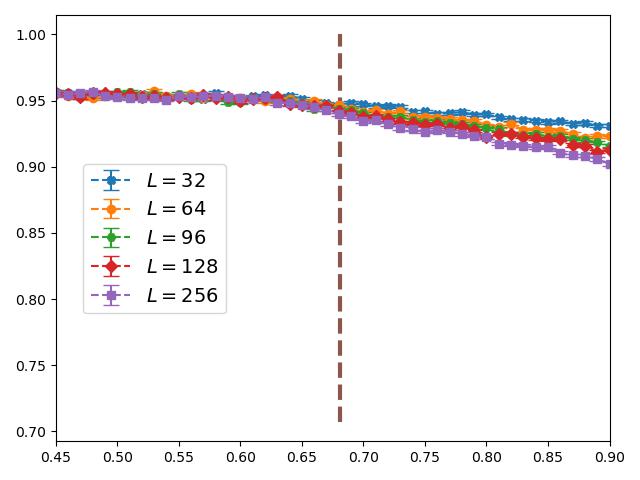}        
        \vskip-0.2cm
        \caption{The smooth single (universal) curve formed by data collapse of various $L$ for the 6-state clock model.
          The vertical dashed line is the expected $T^2_c$.}
        \label{fig62}
\end{figure}

The magnitude $R$ of the NN output vectors as functions of $T$ for various $L$ for the 6-state clock model are shown in fig.~\ref{fig61}.
The figure implies that there are possbily two phase transitions: one before and one after $T\sim 0.6$.
The dashed vertical line is the expected transition temperature $T^2_c$ from LRO to PLRO.
From the figure, one observes that there is a range of $T$ where data of
various $L$ collapse to form a single (universal) curve. In addition, such a
universal curve seems to end at $T^2_c$. This can be considered as a NN
method of estimating $T^2_c$. Based on such an idea,
the NN prediction for $T^2_c$ is found to be 0.67(2)
(see fig.~\ref{fig62}). The obtained $T^2_c$ agrees reasonably well
with the MC result of 0.681 determined in Ref.~\cite{Cha18}.
It should be pointed out that this NN method of calculating $T^{2}_c$ is
not of high precision. Despite this, it leads to a NN prediction of $T^2_c$
with acceptable quality.

\begin{figure}
       \includegraphics[width=0.4\textwidth]{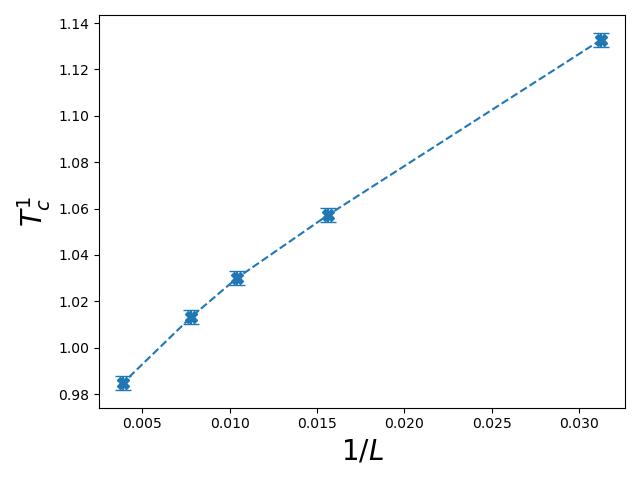}        
        \vskip-0.2cm
        \caption{$T^1_c(L)$ as a function of $1/L$ for the 6-state clock model.}
        \label{fig63}
\end{figure}

After establishing a NN method of estimating $T^2_c$, we turn to the
determination of the transition temperature $T^1_c$ from PLRO to paramagnetic phase.
This transition is similar to that of the 2D classical
$XY$ model. Hence we will use the standard analysis procedure to calculate
$T^1_c$. First of all, one notices that $R$ will take the value of $1/\sqrt{2}\sim 0.70712$
at extremely high temperatures. As a result, for a given $L$, we will consider the intersection between
the related $R$ data (as a function of $T$) and the curve of $2 T/\pi q + 0.70712$ with $q=6$
to be the estimated $T^{1}_c(L)$. With this approach, $T^{1}_c(L)$ as a function of $1/L$
is shown in fig.~\ref{fig63}.

It is anticipated that the $T^1_c(L)$ should fulfill the following ansatz \cite{Nel77,Pal02}
\begin{equation}
T^1_c(L) = T^1_c + \frac{b}{\left(\log\left(L\right)\right)^2},
\end{equation}
where $b$ is some constant. A fit using the data of fig.~\ref{fig63} and above ansatz
leads to $T^1_c = 0.890(5)$. The obtained $T^1_c = 0.890(5)$ agrees
quantitatively with the expected $T^1_c \sim 0.892$.

We would like to point out that analytically the finite-size scaling ansatz for the transition temperature $T_{\text{BKT}}$ of
a BKT phase transition is given by
\begin{equation}
  \label{ABKT}
T_{\text{BKT}}(L) = T_{\text{BKT}} + a \frac{T_{\text{BKT}}}{\left(\log\left(L\right)+c\right)^2}, 
\end{equation}
where $a$ and $c$ are some constants. A fit using this ansatz and the data of fig.~\ref{fig63} leads to $T^1_c = 0.893(15)$
which also matches well with $T^1_c \sim 0.892$.

\subsection{The NN results associated with 8-state clock model}

\begin{figure}
       \includegraphics[width=0.4\textwidth]{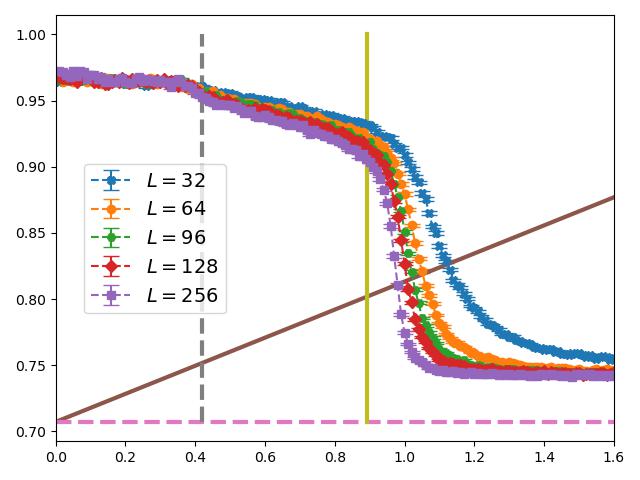}        
        \vskip-0.2cm
        \caption{$R$ as functions of $T$ for various $L$ for the 8-state clock model.
          The dashed horizontal line is 0.70712. The vertical dashed and solid lines
        are the expected transition temperatures $T^{2}_c$ and $T^1_c$. The tilted line is $2 T/(\pi q) + 0.70712$ with $q=8$.}
        \label{fig81}
\end{figure}

\begin{figure}
       \includegraphics[width=0.4\textwidth]{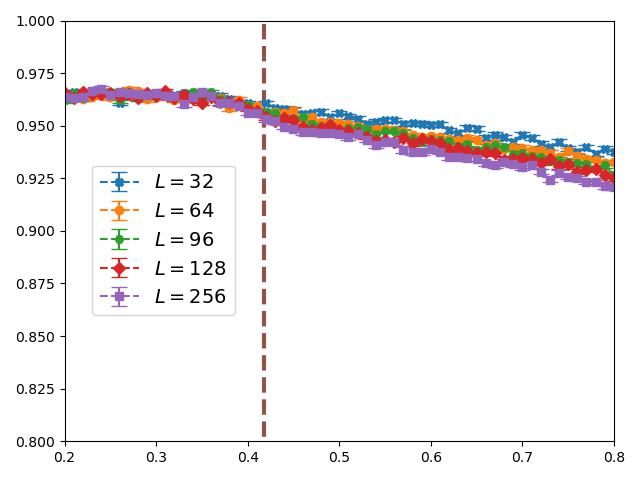}        
        \vskip-0.2cm
        \caption{The smooth single (universal) curve formed by data collapse of various $L$ for the 8-state clock model.
          The vertical dashed line is the expected $T^2_c$.}
        \label{fig82}
\end{figure}

\begin{figure}
       \includegraphics[width=0.4\textwidth]{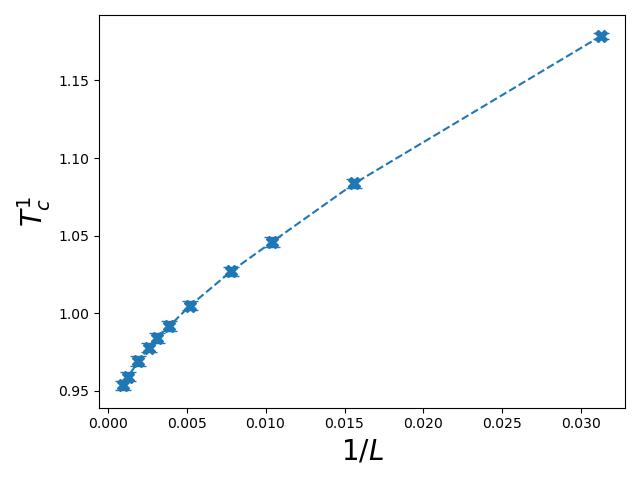}        
        \vskip-0.2cm
        \caption{$T^1_c(L)$ as a function of $1/L$ for the 8-state clock model.}
        \label{fig83}
\end{figure}

The magnitude $R$ of the NN output vectors as functions of $T$ for various $L$ for the 8-state clock model are shown in fig.~\ref{fig81}.
The conventions of the dashed and solid lines in the figure are similar to those used in fig~\ref{fig61}.
Following the same procedures for determining both the transition temperatures $T^2_c$ and $T^1_c$ of the 6-state clock model,
the values of $T^2_c$ and $T^1_c$ for the 8-state clock model are calculated to be 0.40(2) and 0.884(7), respectively, sees figs.~\ref{fig82}
and \ref{fig83}. We would like to point out that the tilted curve in fig.~\ref{fig81} is given by $2T/\left(\pi q\right) +0.70712$ with $q=8$
and $T^1_c = 0.884(7)$ is obtained with data of $192 \le L \le 1024$.
The obtained $T^2_c=0.40(2)$ is in good agreement with  $T^2_c = 0.418$ found in Ref.~\cite{Cha18}. Moreover, the determined
$T^1_c$ of the 8-state clock model matches nicely with the expected $T^1_c\sim 0.892$ as well.

Finally, with a fit using equation~(\ref{ABKT}) and data of fig.~\ref{fig83} ($64 \le L\le 1024$), we arrive at $T^1_c = 0.890(15)$.

\section{Discussions and Conclusions}

In this study, we calculate the transition temperatures $T^1_c$ and $T^2_c$
of the 6- and 8-state clock models using the technique of supervised NN.
In particular, the employed supervised NN has extremely simple architecture,
namely it consists of one input layer, one hidden layer of two neurons, and
one output layer. The supervised NN is trained without any input from the
considered models.

By considering the magnitude $R$ of the NN output vectors as functions of
$T$ and $L$, the values of $T^1_c$ and $T^2_c$ are estimated using semi-experimental
methods. The obtained NN outcomes of the transitions temperatures are in nice
agreement with the corresponding results established in the literature.

Typically, one needs to construct a new NN whenever a new model or a different
system size is considered. The outcomes shown in Refs.~\cite{Tan21,Tse22} and here demonstrate
that a NN with simple infrastructure can be recycled to investigate the phase
transitions of many 3D and 2D models. 

We would like to emphasize the fact that due to the unique features of the employed supervised NN,
there is no system size restriction in our calculations. As a result, outcomes of $L=1024$ can be
reached with ease.

In Refs.~\cite{Bae09,Cha18}, two Binder ratios $U_4$ and $U_m$ are built to detect the the phase transitions associated with
$T^1_c$ and $T^2_c$, respectively. It should be noticed that $U_4$ ($U_m$) cannot be used for studying the phase transition related to
$T^{2}_c$ ($T^1_c$). Particularly, one needs to analytically investigate the model in order to construct the suitable observables to calculate
the transition temperatures. For the present study, one single quantity, namely $R$ can reveal clear signals of both the transitions.
The use of $R$ is natural and does not require any prior investigation of the targeted system(s). This feature can be considered as one
advantage of our NN approach.

For the traditional methods, one can directly use the associated analytic predictions to perform the finite-size scaling analysis
to extract the relevant
physical quantities such as the critical points. For the NN method, such theoretical formulas typically do not exist and one has to rely on
semi-experimental ansatzes to conduct the tasks. In particular, to establish these semi-experimental anstazes requires certain efforts
of investigations, and the employed ansatzes may not be applicable to all cases. From this point of view, the NN method is still in the
developing phase and there is room for improvement.

Finally, we would like to emphasize the fact that similar to the autoencoder and the generative adversarial network
constructed in Ref.~\cite{Tse22}, it is anticipated that the simple supervised NN employed in this study can be directly applied to
study the phase transitions of other models, such as the three-dimensional (3D) $O(3)$ model, the 2D generalized $XY$ model,
the one-dimensional (1D) Bose--Hubbard model, and the 2D $q$-state ferromagnetic Potts model. In other words, it is likely that
one can construct a universal NN that is applicable to investigate the phase transitions of many systems.

\section*{Acknowledgement}\vskip-0.3cm
Partial support from National Science and Technology Council (NSTC) of 
Taiwan (MOST 110-2112-M-003-015 and MOST 111-2112-M-003-011) 
is acknowledged.

\end{document}